# A Solid-state Sub-nm Pore for Single-mer Resolution Sequencing

Jianxin Yang†, Dehua Hu†, Wu Yuan†, Tianle Pan, and Ho-Pui Ho*

Department of Biomedical Engineering, The Chinese University of Hong Kong, Hong Kong SAR, China.

*Corresponding author. Email: aaron.ho@cuhk.edu.hk (H.-P.H.).

†The three authors contributed equally to this work.

## Abstract

Nanopore sequencing accuracy is inherently limited by the quality of data from individual molecular translocation events, requiring advances beyond traditional sequencing-by-synthesis methods. We introduce an oxidized pyramidal sub-nm pore (OPSP) integrated in a three-terminal sensing platform, where the sub-nm silicon pore functions as an electrode for detecting displacement currents across an oxide barrier, induced by counter-ion migration within the electric double layer. This platform achieves sub-1-nm-scale spatial resolution and a signal-to-noise ratio (SNR) up to 15 for biopolymer sequencing, enabling direct identification of individual bases in single-stranded DNA and single amino acids in peptides, with raw-read accuracies exceeding 98.5% and 95.5%, respectively, without consensus-based computational correction. The OPSP demonstrates high acid tolerance, reusability in varied chemical environments, and operational stability for over six months. This work establishes OPSP as a durable, high-accuracy platform for single-mer resolution sequencing, defining a reliable and robust paradigm for next-generation sequencing technologies.

## Introduction

Biological nanopore sequencing has transformed genomics by enabling in situ, label-free analysis of DNA and RNA[1-3]. Commercial platforms, most notably those from Oxford Nanopore Technologies (ONT), use engineered protein pores and motor proteins to achieve scalable sequencing, leveraging ionic current blockades as nucleic acids translocate through the pore[2,4-8]. Beyond nucleic acids, biological nanopores also show potential for long-read protein fingerprinting[9-13]. However, signal detection is inherently context-dependent: each current level reflects a composite signal from an oligonucleotide or peptide segment (typically 4–8 residues), rather than a single nucleotide or amino acid. This convolution results in high raw-read error rates (0.5–2% for DNA; ~13% for proteins) and complicates homopolymer resolution[1,2,4,10,14]. Additionally, the fragility of lipid bilayers, environmental sensitivity of protein pores, and difficulty in tuning pore geometry constrain their operational stability and integration density[10,15-17].

Solid-state nanopores, fabricated in materials such as SiNx, graphene, and $MoS_2$, present an attractive alternative to biological pores, offering enhanced mechanical stability, tunable pore diameters, and potential for high-throughput, array-based sequencing[18-22]. However, conventional solid-state nanopores typically exhibit relatively large pore sizes (>2 nm), short dwell times (<200 μs per base), and low signal-to-noise ratios (SNR <10) in ionic current measurements, which limit their effectiveness for DNA sequencing[23-30]. A central challenge is the reliable fabrication of robust, sub-nanometer pores capable of matching or surpassing the spatial resolution and signal sensitivity of biological nanopores[15].

To overcome the limitations of conventional ionic current sensing, alternative detection modalities with heightened sensitivity to molecular conformations within the nanopore have been explored. Capacitive coupling and tunnelling current measurements, in particular, offer superior SNR for sequencing applications. The capacitive signal, arising from displacement currents generated by transient ion charge imbalances during molecular translocation, enables high sensitivity by probing molecule–pore interactions at atomic length scales and exhibits reduced susceptibility to low-frequency noise[19,31-33].

This work introduces an oxidized pyramidal sub-nm pore (OPSP) platform integrated with a three-terminal sensing configuration, enabling direct sequencing of unmodified single-stranded DNA (ssDNA) and peptides at single-mer resolution. Utilizing a self-limiting wet oxidation process for precise fabrication of 0.55-nm pores and a novel electronic double layer (EDL) displacement current detection scheme, OPSP achieves raw read accuracies exceeding 98.5% for nucleotides and 95.5% for amino acids at the single-mer level. The platform exhibits exceptional chemical stability and operational robustness, with no performance degradation over six months and during a 140-hour continuous operation. These attributes also facilitate real-time monitoring of dynamic biochemical processes, such as pH-triggered peptide release. This work thus bridges the gap between robust solid-state pores and reliable next-generation sequencing technologies.

## Durable oxidized pyramidal sub-nm pore

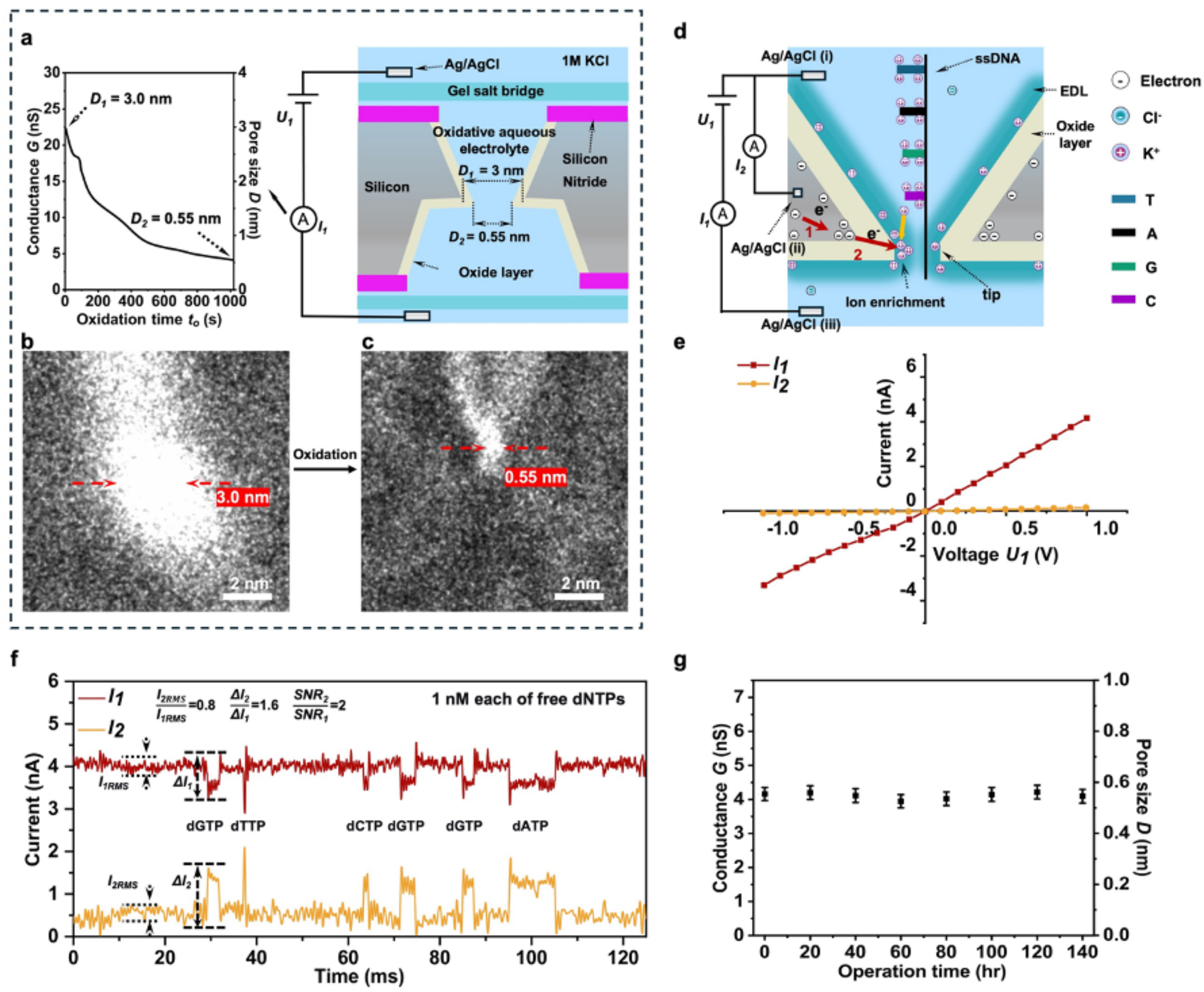

**Figure 1. OPSP fabrication and three-terminal sensing platform. a,** Schematic of an in-situ pore oxidation and size measurement using closed-loop control. The setup features a pore in a flow cell immersed in the solution

of oxidative aqueous electrolyte (30% wt $H_2O_2$ in 1M KCl solution). Two electrodes apply a voltage bias and measure ionic conductance of pores. The initial pore size after wet etching is $D_1$, which shrinks to final pore size $D_2$ after oxidative treatment. This configuration allows for real-time pore size monitoring and the control of $SiO_2$ formation, enabling precise adjustment of the pore size through oxidative shrinkage. The top curve is a representative conductance-time trace showing a monotonic decay as $SiO_2$ grows on the pore wall as the pore size decreases. The curve asymptotically approaches a steady value due to self-limiting oxidation and transport constraints in the narrowed pore. **b and c,** TEM images of a pore before and after oxidation. Fig. b, as-fabricated silicon pore with an estimated size of 3 nm. Fig. c, the same pore after oxidation, showing a 0.55 nm constriction. **d,** Schematic of the three-terminal sensing platform. The translocation of target molecules (e.g., ssDNA) with counterions (e.g., $K^+$) through a pore induces an enrichment of ions in the EDL at the thin oxide layer by ions from the molecule. The transient charge imbalance across the oxide film induces two significant displacement current phenomena: (1) charging of the equivalent oxide-layer capacitance and (2) enhancement of tunneling current through the layer. **e,** The measured $I_1$-$V$ and $I_2$-$V$ curves of the 0.55-nm pore in 1 M KCl solution. $I_1$ and $I_2$ are the measured ionic current and displacement current, respectively. $U_1$ is the applied transmembrane bias. **f,** Current traces of conventional ionic blockade signal ($\Delta I_1$) and displacement signal ($\Delta I_2$) of four free nucleotides (dNTPs) with concentration of 1 nM in 1M KCl solution ($U_1$=1.5 V, pH=7.0). **g,** Longitudinal measurements of conductance ($G$) and corresponding pore size ($D$) of OPSP were taken across eight tests, each conducted after 20 hours of operation. Error bars indicate the standard deviation of conductance $G$ measured from three independent 0.55-nm OPSP samples.

OPSPs are fabricated using wet oxidation combined with closed-loop feedback control, enabling reduction of silicon pore diameters to sub-nanometer scales. The initial pore size is precisely set to 3–5 nm using an electrochemical etch-stop technique[34,35] (see Methods and Supplementary Note 1). A flow-cell electrochemical platform (see Methods) was developed to facilitate self-limiting oxidative narrowing of silicon pores while providing real-time monitoring of ionic conductance. This allows direct measurement of pore size ($D$) from the conductance-time curve ($G$-$t_o$, Figure 1a). Here, the relationship between $G$ and $D$ is described by the equation of $G = \frac{\pi D \sigma}{3.68}\sqrt{\frac{90-\theta}{90}}$ [36,37], where $\sigma$ and $\theta$ are the solution conductivity and the slanted pore wall angle (54.7°). To prevent catalytic decomposition of $H_2O_2$ and bubble intrusion, the electrode in 1 M KCl was isolated from the oxidative aqueous electrolyte using a gel salt bridge, ensuring reproducible electrochemical measurements and a stable baseline. The oxidation time required to achieve the target sub-nanometer pore size ($D_2$) depends on the initial etched pore size ($D_1$). For instance, starting from 3.0 nm (Figure 1b), the pore was narrowed to 0.55 nm after about 1000 seconds of oxidation (Figure 1c). The oxidation process is self-limiting, exhibiting a progressively decreasing pore-narrowing rate from 0.906 nm/min to 0.012 nm/min. This behaviour enables precise control of the final pore size and the oxide layer thickness to approximately 1.25 nm (Figure 1a and Supplementary Note 2).

To enhance OPSP sensitivity for sequencing applications, we developed a three-terminal pore sensing technique (Figure 1d). This system comprises an ionic current ($I_1$) circuit between electrodes i and ii, and a displacement current ($I_2$) circuit between electrodes i and iii (see Methods and Supplementary Note 3). The approach effectively decouples the transmembrane bias ($U_1$) from the displacement current. Upon translocation of ssDNA and associated counterions (e.g., $K^+$) through the pore, a pronounced transient disturbance in the ionic current occurs. The conventional ionic blockade signal ($\Delta I_1$) arises from physical obstruction of the pore by DNA, which increases local resistance[3]; however, the influx of excess counterions accompanying DNA enhances local conductivity, partially offsetting this effect[38]. In contrast, the displacement current signal ($\Delta I_2$) does not depend on pore blockage by DNA. Instead, it originates from charging of the oxide layer's equivalent capacitance and enhancement of the tunneling current, and is positively correlated with ion enrichment in the EDL at the pore tip induced by translocation events (Figure 1d). Ion enrichment results from both migration of pre-existing ions within the pore and introduction of additional counterions. Consequently, the amplitude of $\Delta I_2$ exceeds that of $\Delta I_1$, while the baseline ionic current $I_1$ is significantly higher than the displacement current $I_2$ (Figure 1e). In addition, since the root-mean-square (RMS) circuit noise is proportional to the capacitive feedback in the circuit, $I_{2,RMS}$ is lower than $I_{1,RMS}$ (Supplementary Note 4)[28].

To assess signal-to-noise ratio (SNR) and resolution, we tested four types of free dNTPs (Supplementary Note 5). Simultaneous recordings of $I_1$ (red) and $I_2$ (yellow) from an equimolar dNTP mixture (1:1:1:1) revealed distinct signal signatures for each nucleotide (Figure 1f). The displacement current exhibited molecular dwell times identical to those observed in the ionic current, while achieving an $SNR_2$ of up to 6, approximately twice that of the ionic current ($SNR_1$), for free nucleotides.

Notably, OPSPs maintain stable performance with no observable degradation after more than 140 hours of cumulative sequencing and preserves full functionality over six months of operation, demonstrating exceptional long-term stability (Figure 1g). Additionally, as the EDL is confined to the pore wall in 1M KCl, the displacement current signal is exclusively generated by molecular interactions with the local EDL at the pore tip. This inherent localization confers a spatial resolution for displacement current sensing that surpasses that of conventional ionic current measurements.

## Single-base resolution sequencing of unmodified ssDNA

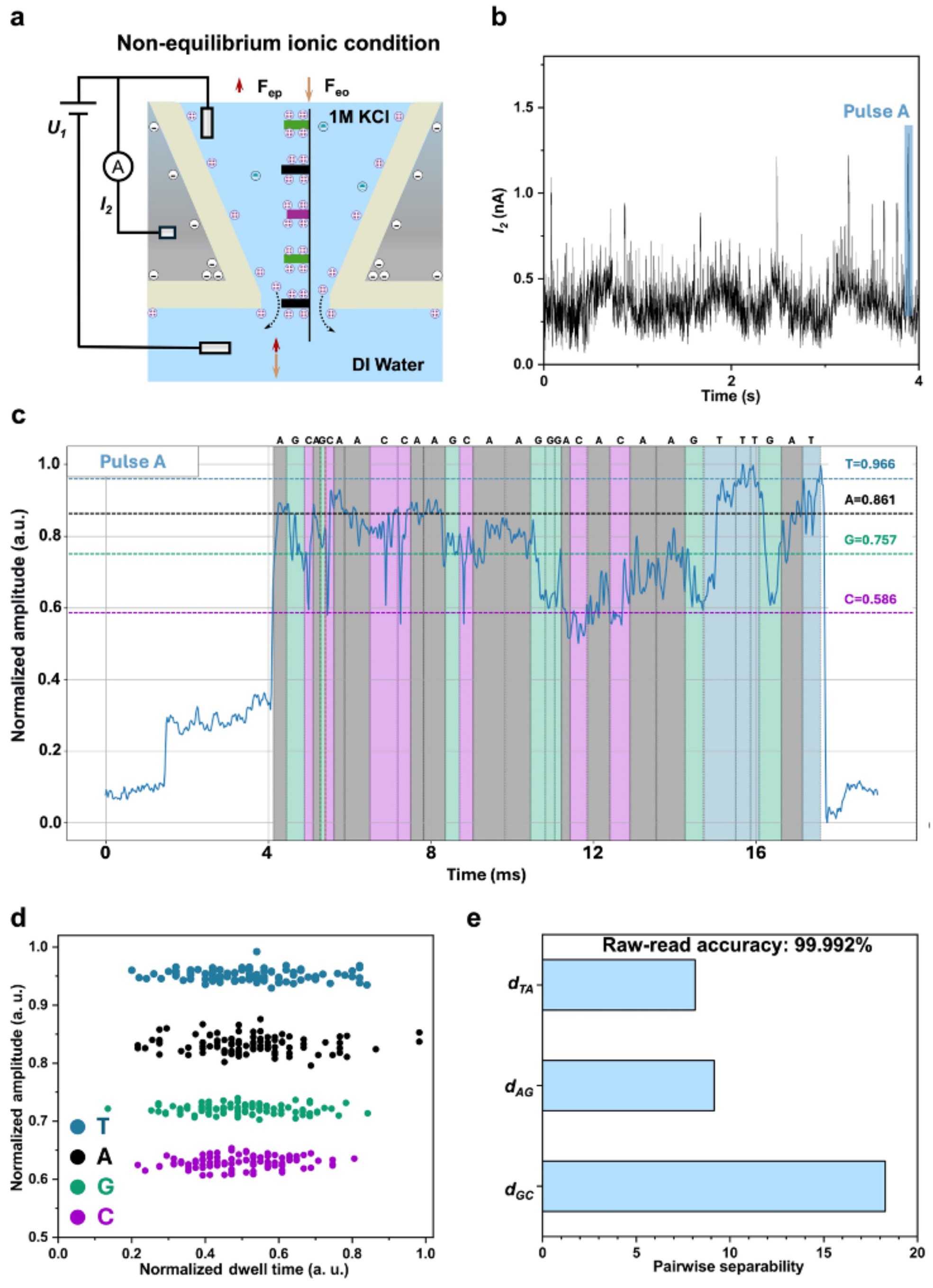

**Figure 2. Sequencing under non-equilibrium ionic concentrations across the pore. a,** Schematic illustration of sequencing of ssDNA under a non-equilibrium ionic condition (1 M KCl in pore with deionized water outside) and an applied voltage $U_1$ of 1.5V. The red arrow ($F_{ep}$) and brown arrow ($F_{eo}$) denote the electrophoretic and electroosmotic forces exerting on the ssDNA molecules, respectively, which $F_{eo}$ substantially stronger than $F_{ep}$. **b,** Raw trace of the displacement current ($I_2$) over a 4-second time window. **c,** A representative translocation current trace (highlighted in blue as Pulse A in II). Output signal decoded into discrete events for individual bases. Dwell regions are colored: T (cyan), A (black), G (teal), and C (purple). The four dashed lines indicate the average normalized current levels for the four nucleotides. **d,** Distributions of normalized current amplitude versus normalized dwell time for the four bases, measured across 20 ssDNA translocation events. Maximum current

amplitude and longest dwell time of discrete base events were used for normalization (set to 1). **e,** Bar plot of pairwise separation metrics. The plot shows the pairwise separability of bases with closely normalized current levels ($d_{TA}$, $d_{AG}$, and $d_{GC}$) and the calculated raw-read accuracy. Values are derived from the mean and standard deviation of the 640 discrete base events.

The three-terminal OPSP platform was initially employed to sequence an unmodified 32-base ssDNA (see Methods). Two ionic conditions were tested: equilibrium and non-equilibrium concentrations across the pore. Under equilibrium ionic conditions, ssDNA translocate through the pore via combined electrophoretic and electroosmotic forces, generating characteristic displacement current traces (Supplementary Note 6). Conversely, a higher ionic concentration inside the pore than outside induces concentration polarization at the pore orifice, substantially increasing local electroosmotic flow at the pore tip[39] (Figure 2a and b). The measured average dwell times of the 32-base ssDNA under the non-equilibrium ionic condition were approximately 20 ms (0.65 ms for average per-base dwell time), while the corresponding pulse amplitudes remained comparable in magnitude, with a SNR consistently exceeding 8 (Figure 2c). In addition, this enhanced flow governs ssDNA translocation dynamics and stabilizes signal outputs, as evidenced by a reduction in per-base dwell time coefficient of variation (CV) from 2.41 to 0.85 (Figure 2d; Supplementary Note 6).

To decode corresponding nucleotide sequence from the ssDNA pulse signal traces, we developed a base-calling pipeline. The raw full pulse signals were first normalized in both signal amplitude (Figure 2c). Then, segmented along the time axis into individual events according to the number of bases (e.g., 32), and the current amplitude axis was divided into four distinct regions for corresponding to four type of bases. Based on a known reference sequence, the current pulse levels were calibrated such that the discrete events originating from the same base were assigned to the same amplitude region as consistently as possible. This was achieved through adjustment of the dwell times for every discrete base events.

The error rate for base discrimination was estimated based on the assumption that the signal amplitudes for individual amino acids and nucleotides follow Gaussian distributions, using pairwise separability ($d$) analysis. Here, the pairwise separability (i.e., between bases T and A), was calculated as $d_{TA} = \frac{|\mu_T - \mu_A|}{\sqrt{\frac{(\sigma_T{}^2 + \sigma_A{}^2)}{2}}}$ [40], where $\mu_T$, $\mu_A$, $\sigma_T$, and $\sigma_A$ are the measured average current levels and standard deviations for base T and base A, respectively (Figure 2 c and d). The normalized pulse amplitude versus normalized dwell time distribution for these individual events identified the minimum pairwise separability $d_{min}$ of 8.14 (Figure 2 e). As random misidentification typically occurs for a limited number of reads[41], the ultimate raw-read error rate $P_{error}$ was determined by the identifying the minimum separability of $N\left[1 - \phi\left(\frac{d_{min}}{2}\right)\right]$[42], where $d_{min}$ and $\phi$ denote the minimum pairwise d-prime value and the standard normal cumulative distribution function, respectively. $N$ represents the number of potential error paths, which is used to estimate the total error rate. It is determined as 3 for the 4 nucleotide bases and to 19 for the 20 amino acids. This revealed a raw-read error rate $P_{error}$ less than 0.01% under non-equilibrium ionic condition. This result proves that the OPSP platform can effectively single-base resolve the four bases in short-read ssDNA translocations without the assistance of computational model.

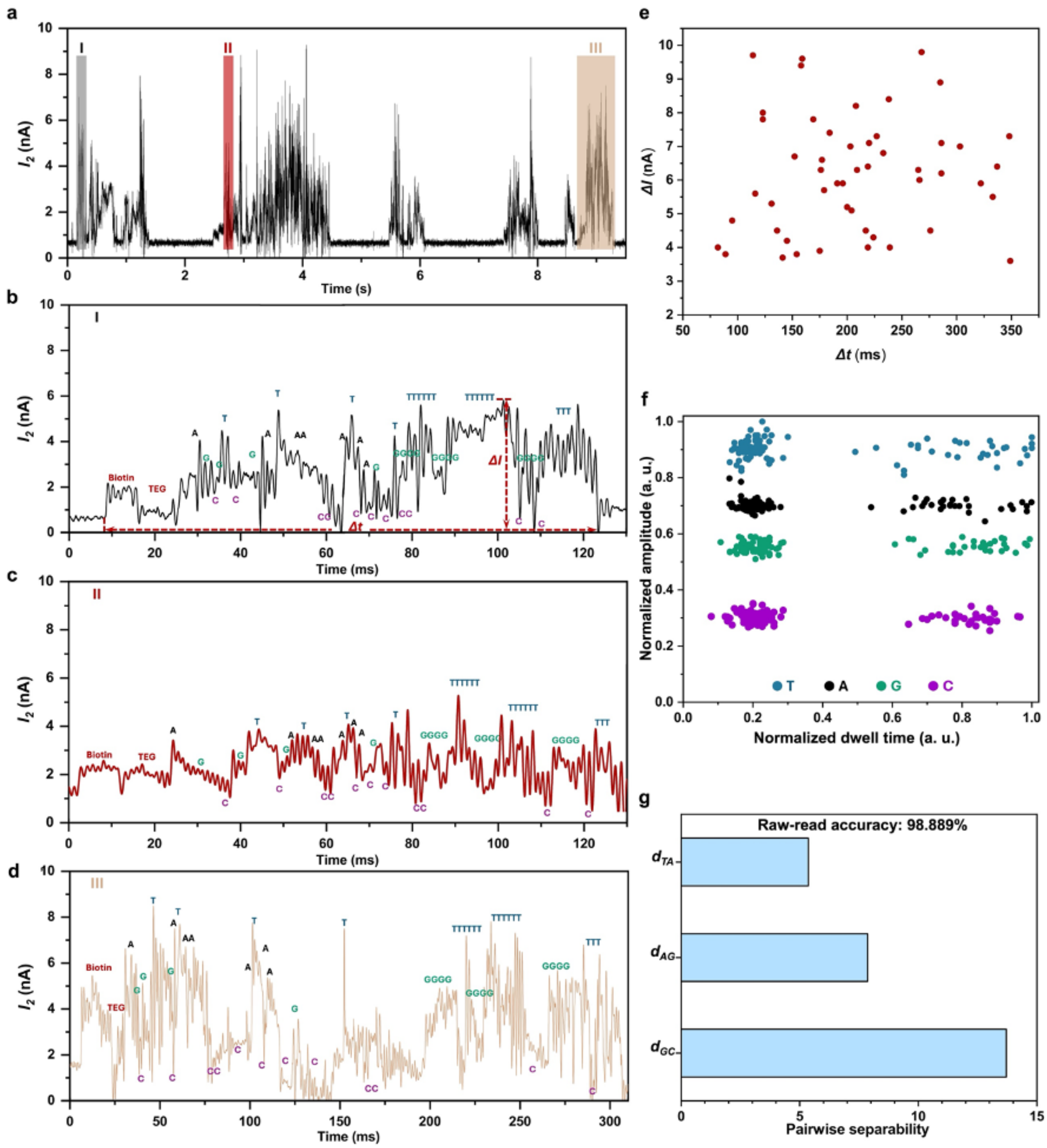

**Figure 3. Sequencing of ssDNA. a,** Raw trace of the displacement current ($I_2$) over a 10-second time window for sequencing a flexible ssDNA with number of bases of 53 under non-equilibrium ionic condition by OPSP with $U_1$ of 1.5V. **b-d,** Representative current pulses of the ssDNA, corresponding to the segments highlighted in gray (Part I), red (Part II), and beige (Part III) in Fig. a, respectively. The predicted nucleotide sequence and the 5'-end modification (Biotin-PEG) are annotated alongside the current levels. The raw data for the three pulses were processed with a band-stop Bessel filter (7-15 Hz) to suppress special frequency noise induced by flexible DNA perturbations. $\Delta I$ and $\Delta t$ in Fig. b denotes the current amplitude and dwell time of the translocation pulse, respectively. **e,** Two-dimensional scatter plot of the pulse amplitude ($\Delta I$) versus dwell time ($\Delta t$) for 100 complete ssDNA translocation events measured under non-equilibrium ionic condition in the OPSP. **f,** Distributions of normalized current amplitude versus normalized dwell time for the four-type nucleotides of T (cyan), A (black), G (teal), and C (purple), measured from 20 ssDNA translocation events in the OPSP. The maximum current amplitude and the longest dwell time among the 52 bases in each pulse were regarded as the normalization benchmarks (set to 1) for the two respective features of all other bases. **g,** Bar plot of pairwise separation metrics.

The plot displays the pairwise separability of bases with closed normalized current level and calculated raw-read accuracy.

53-base ssDNA in 1 M KCl solution behaves as a flexible polymer chain[43]. In this study, the OPSP platform was used to test the 5′-Biotin-TEG-53-base ssDNA-3′ (see Methods) to verify the capability for sequencing flexible ssDNA molecules. Under non-equilibrium ionic condition with the applied voltage $U_1$ of 1.5 V, stable current pulses were observed, corresponding to an average ssDNA translocation frequency of approximately 1 $s^{-1}$ (Figure 3a). After applying band-stop Bessel filtering, three representative pulses exhibited identical sequence patterns (Figure 3b-d). Representative current pulses showed dwell times on the order of 100 ms (4 ms for average per-base dwell time) and a SNR significantly up to 15 (Figure 3e).

Normalization and segmentation of these pulses by dwell time and amplitude confirmed that each base occupies a distinct cluster in the feature space (Figure 3f). This was further confirmed by the consistent pulse levels of base T in both copolymer (e.g., ATAA) and homopolymer (e.g., TTTTTT) contexts (Figure 3b-d). In addition, the order of normalized pulse amplitudes for the four-type nucleotides was T, A, G, and C, consistent with the trend observed for the 32-base ssDNA. The minimum pairwise separability $d_{min}$ was 5.36 (between base T and A), corresponding to a raw-read error rate $P_{error}$ of 1.111% (Figure 3g). These results support the interpretation that the displacement current signal reflects the efficiency with which each base contacted with OPSP tip mobilizes $K^+$ ions. The signal is strongest when the negative charge is localized and sterically accessible (as in bases T and A), while the weaker signals for G and particularly C arise from electrostatic repulsion between their partially or fully negatively charged nucleobases and the phosphate backbone[44,45], which disperses the local electric field and reduces net cation flux[46]. Furthermore, consistency of signal order was also confirmed in long-read of ssDNA (e.g., 7249 base, see Methods and Supplementary Note 7).

**Protein sequencing at single-amino-acid resolution**

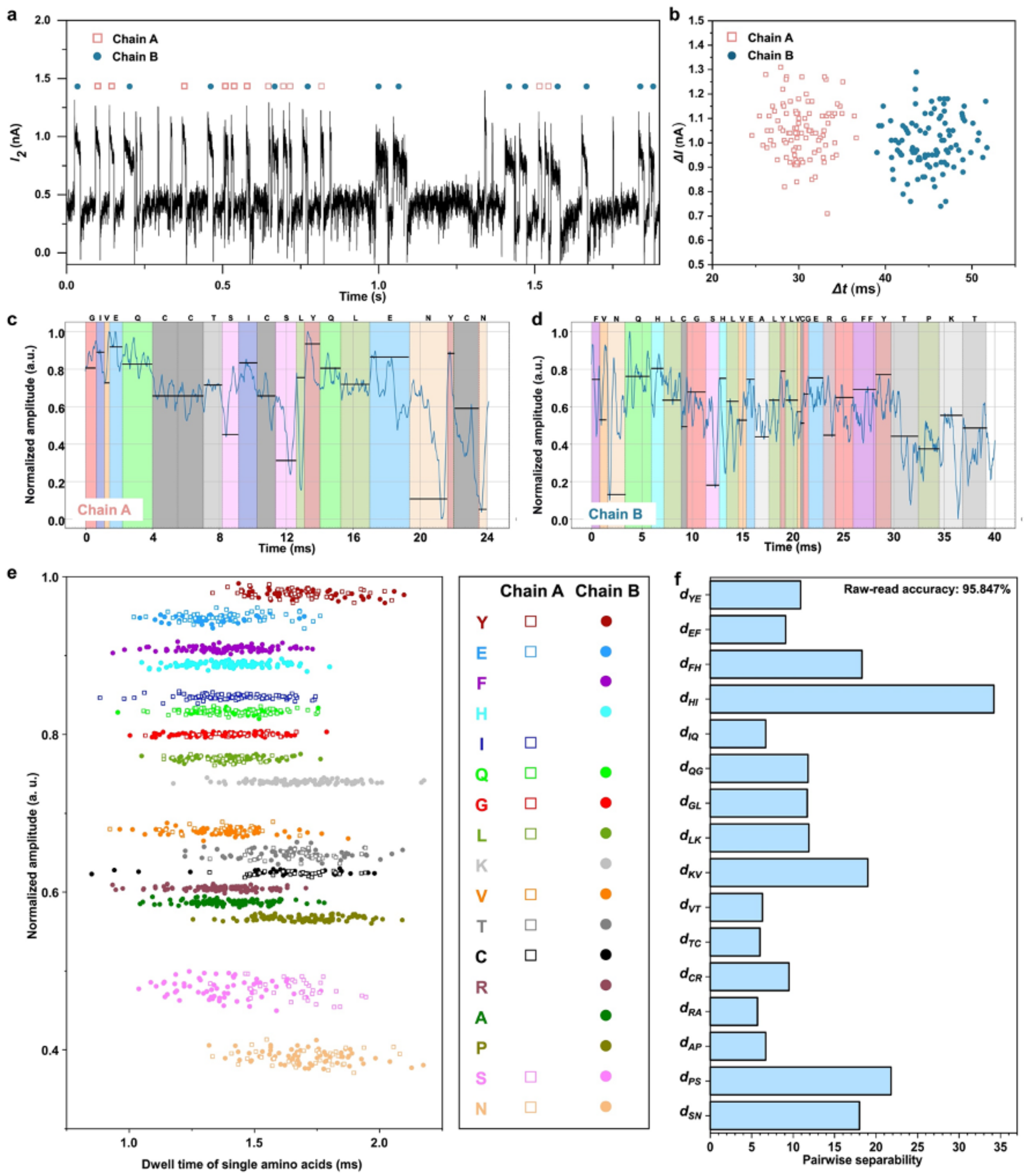

**Figure 4. Sequencing the constituent peptides of insulin. a and b,** Representative current trace (b) and distribution of current amplitude ($\Delta I$) versus dwell time ($\Delta t$) (c) for 200 translocation events measured using the OPSP under the non-equilibrium ionic condition with $U_1$ of 1.5V. The signal traces were obtained after a 100nM insulin sample had been incubated to dissociate the molecules into its constituent peptides (Chain A and Chain B). Translocation events corresponding to Chain A and Chain B are indicated by hollow squares (pink) and solid circles (cyan), respectively. **c and d,** Representative current pulses corresponding to Chain A (c) and Chain B (d). The predicted amino acid sequence for each peptide is annotated alongside the dwell region with different colours. Thin black lines indicate the average normalized current levels for each residue. **e,** Distributions of normalized current amplitude versus dwell time for each amino acid, derived from 10 Chain A (hollow squares) and 10 Chain B (solid circles) translocation events. For each translocation pulse, the current amplitude of every amino acid was normalized to the maximum current amplitude observed within that same pulse, which was set to 1. **f,** Bar plot of pairwise separation metrics. The plot displays the pairwise separability of amino acids with closed normalized current levels and calculated raw-read accuracy.

A benchmark protein (e.g., Insulin) was hydrolysed into peptide chains following incubation with urea and beta-mercaptoethanol (βME) (see Methods). We further sequenced two constituent peptides of Chain A (21 amino acids) and B (30 amino acids) to demonstrate the single-amino-acid resolution of the OPSP platform for protein sequencing. Under the non-equilibrium ionic condition, representative displacement current signals generated from the three-terminal platform clearly revealed distinct pulse signatures for Chain A (pink hollow squares) and Chain B (cyan solid circles) (Figure 4a). The observed ratio of these characteristic signal events was approximately 1:1, consistent with the fact that insulin digestion produces equimolar amounts of Chains A and B.

The current pulses of two peptides exhibited different dwell-time distributions (i.e., 30 ms VS 45 ms), which were significantly correlated with the number of amino acids in each peptide (Figure 4b). Identical peptide chains produced current pulses with highly reproducible sequence patterns, enabling real-time sequencing at single-amino-acid resolution (Figure 4c and d). Notably, compared to the dwell times of ssDNAs with similar length (e.g., 32-base ssDNA), that of the peptides showed greater uniformity, with a CV of approximately 0.3 per amino acid (Figure 4e). After normalization and segmentation of the pulse amplitudes, the data confirmed that each amino acid formed a unique cluster in feature space. This established a minimum pairwise separability ($d_{min}$) of 5.7 between amino acids R and A, which corresponds to a raw-read error rate ($P_{error}$) of 4.153% for clear discrimination (Figure 4f). The direct sequencing capability at single-amino-acid resolution stems from the significant displacement of EDL ions induced by the direct contact between the peptide and the pore tip. In a control study using larger pores (e.g., 1.1 nm), the sequencing resolution deteriorated, with the raw-read accuracy drastically dropping to 23% (Supplementary Note 8).

**Peptide chain sequencing under acidic conditions**

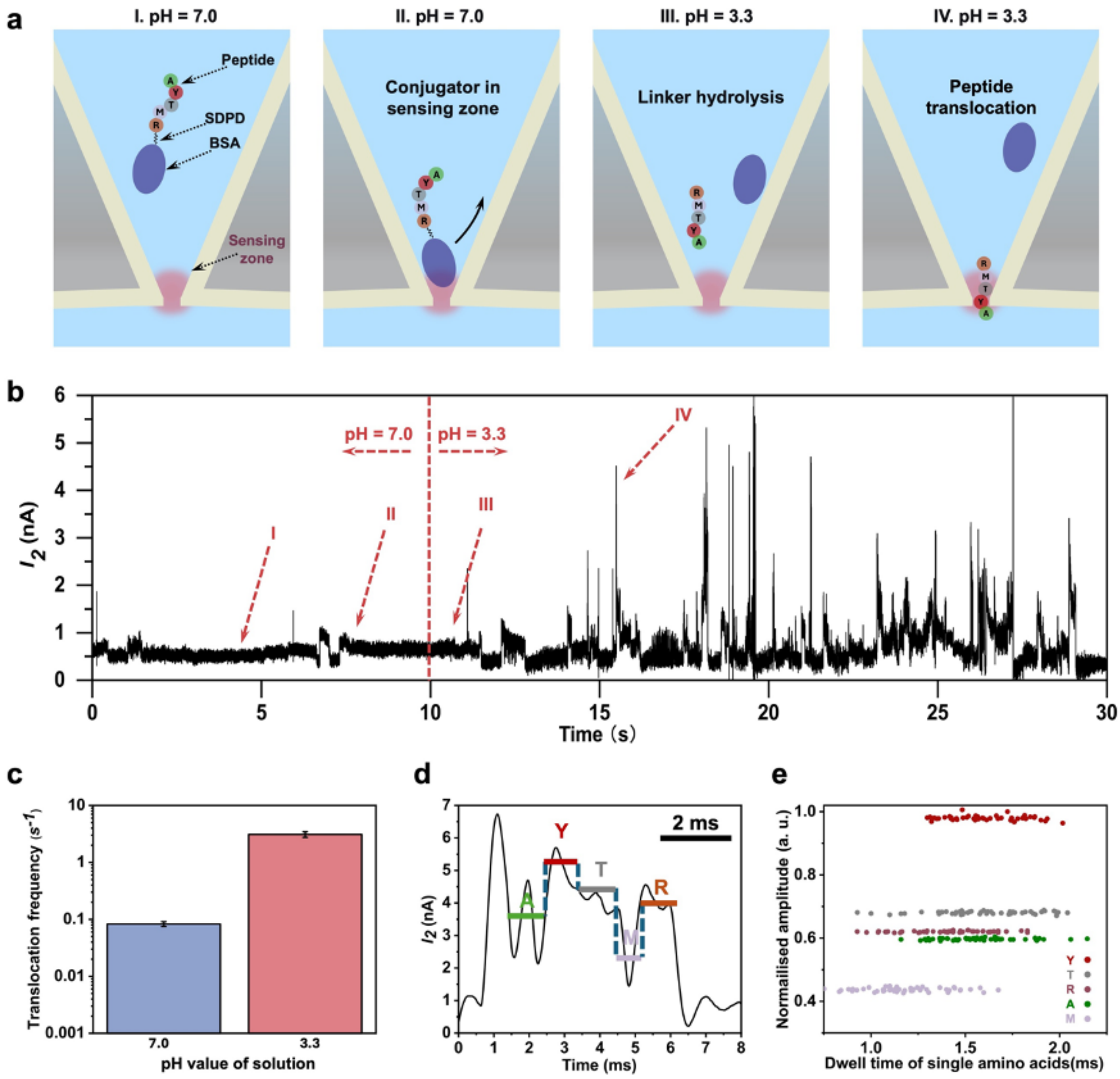

**Figure 5. Real-time monitoring of peptide chain release. a and b,** Schematic illustrations (a) and displacement current trace (b) for real-time monitoring BSA conjugated 5-amino acid peptide with concentration of 1 nM before and after peptide release from BSA. The experiment began with $U_I$ of 1.5V, at pH of 7.0, where outside molecules (I) were driven into the sensing region (II), while BSA is too large to go through the pore. The medium was then changed to pH of 3.3 (III) to break the BSA conjugation. Subsequently, the free peptide chain translocated through the pore (IV) with the translocation of each amino acid reflected in the recorded current pulse signals. **c,** Translocation pulse frequency for the peptide chains at pH of 7.0 (blue) and 3.3 (red). Error bars represent the standard deviations of the frequency, calculated from the first 50 translocation pulses in the recorded current traces. **d,** Representative translocation current pulses corresponding to the peptides. The predicted amino acid sequence for each peptide is annotated alongside the current levels. **e,** Distributions of normalized current amplitude versus dwell time for each amino acid, derived from 20 5-peptide-chain translocation events. For each translocation pulse, the current amplitude of each amino acid was normalized to the maximum current amplitude observed within that same pulse, which was set to 1.

The acid tolerance of the OPSP sequencing platform enables real-time monitoring of the release of specific peptide sequences, such as the acid-triggered cleavage of a short peptide from a BSA-conjugated 5-amino-acid peptide (see Methods). Here, the peptide was crosslinked to BSA via N-succinimidyl-3-(2-pyridyldithio)propionate (SPDP). SPDP remains stable under neutral conditions (pH = 7.0), and thus the conjugate rarely produces translocation events. The current trace only records the baseline (I in Figure 5a and b) and occasional oscillatory pulses as the conjugate enters the sensing region (II in Figure 5a and b). Acidification to pH of 3.3 triggered the rapid dissociation of SPDP and the subsequent release of the peptide (III in Figure 5a and b). This process yields a current trace rich in peptide translocation pulses (IV in Figure 5a and b).

Comparison of current traces before and after acidification reveals a 20-increase in peptide translocation event frequency (0.08 $s^{-1}$ VS 1.62 $s^{-1}$) under acidic conditions (Figure 5c). The pulses in the displacement current trace correspond to the expected peptide sequence (Figure 5d). Following normalization and event segmentation, the current pulse signals for each amino acid exhibit unique and reproducible distributions in the amplitude level with raw-read error rate $P_{error}$ of 1.713%, demonstrating single-amino-acid resolution (Figure 5e). Notably, the normalized amplitude of each amino acid component closely matches that obtained from insulin peptide sequencing (Figure 5e), indicating the consistent ability of the OPSP platform to resolve different protein chains.

## Discussion

Our work establishes OPSP as a robust, scalable platform for single-mer resolution sequencing of nucleic acids and proteins. Integrating a three-terminal configuration to capture displacement currents enables sub-1-nm-scale spatial resolution and a SNR to 15. Pulse signals arise from the migration of pre-existing ions and the introduction of counterions induced by translocation events. This process enriches ions within the EDL at the pore tip, charging the nanometre-scale oxide capacitor and enhancing tunneling current. Owing to its exceptional sensitivity to highly localized ionic perturbations from a single molecule within the pore, this mechanism enables single-mer resolution. This allows direct identification of individual nucleotides and amino acids, achieving raw-read accuracies over 98.5% and 95.5%, respectively, without computational correction. Exceptional acid tolerance supports real-time monitoring of dynamic processes such as pH-triggered peptide release. OPSPs show no performance degradation after 140 hours of continuous use and remains stable for over six months, underscoring its reliability and robustness for diverse sequencing applications.

Despite these advances, challenges remain. Signal stability, enabled by OPSP's solid-state design, is essential for high-fidelity single-mer identification, yet high-throughput decoding of ultralong reads will require further integration of machine learning–assisted signal interpretation into base-calling. Currently, the self-limiting oxidation process for fabricating sub-nm silicon pores achieves a yield rate of approximately 30%. Specifically, batch processing a 2-inch wafer typically produces 15–20 functional OPSP platforms out of 60 fabricated samples, each with a consistent pore size of 0.55 nm. The primary limitation in yield arises from the manual termination of the high temperature pre-etch process, which poses a risk of over-etching. To improve yield, future research should prioritize developing an automated, closed-loop controlled pre-etch technique. Additionally, efforts should be directed toward adapting algorithms to analyse the unique displacement current signatures associated with post-translational modifications. With the implementation of multi-pore arrays and machine learning–based signal decoding, this technology promises to become a foundational tool in proteomics and genomic medicine, significantly increasing the sequencing accuracy and reducing costs through arrayed OPSP platforms.

## Methods

*Electrochemical etch-stop technique for controlling original pore size.* A double-side polished silicon wafer was initially coated with 200-nm-thick silicon nitride layers on both surfaces using plasma-enhanced chemical vapor deposition (Oxford Instruments plc). Photolithography (SUSS MicroTec SE) was subsequently used to pattern the silicon nitride layers on both sides, followed by a 20-minute gaseous plasma treatment using a mixture of carbon tetrafluoride ($CF_4$, 100 sccm) and argon (Ar, 5 sccm) to transfer the square patterns. The front-side pattern had a side length of 30 μm, and the back-side pattern measured 800 μm. An inverted square-base funnel-shaped structure with a depth of approximately 22 μm was then fabricated on the front side by etching in a KOH solution at 80.0 °C. Next, a pre-etching process was conducted. First, the substrate was etched in KOH at 65.0 °C until the remaining silicon thickness reached approximately 100 nm, monitored in real-time by tracking the transmission spectral peak at 480 nm using a wideband blue light source. Finally, a perforation process was implemented to precisely control the initial pore size $D_I$. An electrical circuit was set up to apply a bias of 800 mV and monitor the etching current in real time during KOH etching at 22.0 °C. Perforation was identified by a characteristic rapid increase in the etching current signal. A current ramp with a slope of 0.3 nA/sec was used as the symbol signal to remove the pore sample from the etchant. The initial pore sizes $D_I$ corresponding to the slope threshold exhibited a distribution of 4.0 ± 1.0 nm[34,35] (see Supplementary Note 1 for details).

*Self-limiting wet-oxidation process.* The setup comprises a dual-chamber flow cell, Ag/AgCl electrodes, and a circuit for applying a forward bias $U_I$ of 800 mV and measuring the ionic current signal $I_I$. The two chambers are separated by a membrane supporting the pore chip. The Ag/AgCl electrodes are positioned at either end of the flow cell and connected to dedicated side chambers of the main compartment via gel salt bridges for stable and low-noise measurements. The oxidant used was 30 wt% $H_2O_2$ in 1 M KCl solution. After injecting the oxidant into the main chamber, the pore size evolution was monitored in real time through the conductance-time ($G$-$t_o$) curve.

*Three-terminal pore sensing platform configuration.* The experimental setup consisted of four primary components, including three Ag/AgCl electrodes (i, ii, and iii), an OPSP sample, front and back side flow cells, and a current signal detection module (see Supplementary Note 2). Electrodes i and iii were connected to the electrolytes within the front and back side cells, respectively. The Ag/AgCl electrode ii was formed by directly coating and curing a commercial paste on a pre-treated silicon substrate (HF-etched for 15 seconds), establishing an ohmic contact, and was subsequently encapsulated with a silicone film to electrically insulate it from the solution. The signal detection module utilized an ultra-low bias operational amplifier circuit (OPA128, Burr-Brown) to amplify the displacement current pulses. This circuit achieved a sensitivity of 1.1 V/nA and exhibited a low-level RMS circuit noise of 0.4 pA when sampled at 50 kHz.

*Preparation of Biomolecular Samples.* The proteins and short ssDNA molecules used in this study included a 32-base DNA (sequence: AGC AGC AAC CAA GCA AGG GAC ACA AGT TTG AT), a 5′-Biotin-TEG-53-base ssDNA-3′ (sequence: AGC GTC GAT AAC CAT ACA CGC TCC GGG GTT TTT TGG GGT TTT TTC GGG GCT TT), and Insulin, all purchased from Thermo Fisher Scientific, Inc. The 7249-base long ssDNA (M13mp18) was obtained from New England Biolabs, Inc. A BSA conjugate with a 5-amino-acid peptide (sequence: AYTMR) via SPDP linkage was sourced from GenScript, Inc. All molecular samples were dispersed in a 1 M KCl solution. The pH of the samples was adjusted by the addition of hydrochloric acid and precisely calibrated using a pH meter (model PH-3C, YK Scientific Instrument Co., Ltd.).

*Reductive denaturation of insulin into chains A and B.* 4 μL Insulin (1 μg/μL) was denatured by the sequential addition of an equal volume of βME followed by three volumes of 8 M urea. The mixture was incubated at 55–70 °C for 60 min (or at 22 °C for 120 min) to facilitate complete reduction of disulfide bonds and disruption of the native structure. The resulting peptide chains were then diluted approximately 343 times with 1 M KCl to a final concentration of 100 nM for sequencing using a sub-nanometer pore.

## References

1 Deamer, D., Akeson, M. & Branton, D. Three decades of nanopore sequencing. *Nature Biotechnology* **34**, 518-524 (2016).

2 Jain, M., Olsen, H. E., Paten, B. & Akeson, M. The Oxford Nanopore MinION: delivery of nanopore sequencing to the genomics community. *Genome Biology* **17**, 239 (2016).

3 Wang, Y., Zhao, Y., Bollas, A., Wang, Y. & Au, K. F. Nanopore sequencing technology, bioinformatics and applications. *Nature Biotechnology* **39**, 1348-1365 (2021).

4 Laszlo, A. H. *et al.* Decoding long nanopore sequencing reads of natural DNA. *Nature Biotechnology* **32**, 829-833 (2014).

5 Rang, F. J., Kloosterman, W. P. & de Ridder, J. From squiggle to basepair: computational approaches for improving nanopore sequencing read accuracy. *Genome Biology* **19**, 90 (2018).

6 Cherf, G. M. *et al.* Automated forward and reverse ratcheting of DNA in a nanopore at 5-Å precision. *Nature Biotechnology* **30**, 344-348 (2012).

7 Lieberman, K. R. *et al.* Processive replication of single DNA molecules in a nanopore catalyzed by phi29 DNA polymerase. *Journal of the American Chemical Society* **132**, 17961-17972 (2010).

8 Garalde, D. R. *et al.* Highly parallel direct RNA sequencing on an array of nanopores. *Nature Methods* **15**, 201-206 (2018).

9 Ouldali, H. *et al.* Electrical recognition of the twenty proteinogenic amino acids using an aerolysin nanopore. *Nature Biotechnology* **38**, 176-181 (2020).

10 Brinkerhoff, H., Kang, A. S., Liu, J., Aksimentiev, A. & Dekker, C. Multiple rereads of single proteins at single–amino acid resolution using nanopores. *Science* **374**, 1509-1513 (2021).

11 Zhang, S. *et al.* Bottom-up fabrication of a proteasome–nanopore that unravels and processes single proteins. *Nature Chemistry* **13**, 1192-1199 (2021).

12 Restrepo-Pérez, L., Joo, C. & Dekker, C. Paving the way to single-molecule protein sequencing. *Nature Nanotechnology* **13**, 786-796 (2018).

13 Motone, K. *et al.* Multi-pass, single-molecule nanopore reading of long protein strands. *Nature* **633**, 662-669 (2024).

14 Lin, X. *et al.* High accuracy meets high throughput for near full-length 16S ribosomal RNA amplicon sequencing on the Nanopore platform. *PNAS Nexus* **3**, pgae411 (2024).

15 Branton, D. *et al.* The potential and challenges of nanopore sequencing. *Nature Biotechnology* **26**, 1146-1153 (2008).

16 Kasianowicz, J. J., Brandin, E., Branton, D. & Deamer, D. W. Characterization of individual polynucleotide molecules using a membrane channel. *Proceedings of the National Academy of Sciences* **93**, 13770-13773 (1996).

17 Stoddart, D., Heron, A. J., Mikhailova, E., Maglia, G. & Bayley, H. Single-nucleotide discrimination in immobilized DNA oligonucleotides with a biological nanopore. *Proceedings of the National Academy of Sciences* **106**, 7702-7707 (2009).

18 Dekker, C. Solid-state nanopores. *Nature Nanotechnology* **2**, 209-215 (2007).

19 Feng, J. *et al.* Identification of single nucleotides in MoS2 nanopores. *Nature Nanotechnology* **10**, 1070-1076 (2015).

20 Merchant, C. DNA translocation through graphene nanopores. *Biophysical Journal* **100**, 521a (2011).

21 Kwok, H., Briggs, K. & Tabard-Cossa, V. Nanopore fabrication by controlled dielectric breakdown. *PloS One* **9**, e92880 (2014).

22 Waugh, M. *et al.* Solid-state nanopore fabrication by automated controlled breakdown. *Nature Protocols* **15**, 122-143 (2020).

23 Schneider, G. F. *et al.* DNA translocation through graphene nanopores. *Nano Letters* **10**, 3163-3167 (2010).

24 Carlsen, A. T., Zahid, O. K., Ruzicka, J., Taylor, E. W. & Hall, A. R. Interpreting the conductance blockades of DNA translocations through solid-state nanopores. *ACS Nano* **8**, 4754-4760 (2014).

25 Smeets, R. M. *et al.* Salt dependence of ion transport and DNA translocation through solid-state nanopores. *Nano Letters* **6**, 89-95 (2006).

26 Chou, Y.-C. *et al.* Coupled nanopores for single-molecule detection. *Nature Nanotechnology* **19**, 1686-1692 (2024).

27 Leitao, S. M. *et al.* Spatially multiplexed single-molecule translocations through a nanopore at controlled speeds. *Nature Nanotechnology* **18**, 1078-1084 (2023).

28 Rosenstein, J. K., Wanunu, M., Merchant, C. A., Drndic, M. & Shepard, K. L. Integrated nanopore sensing platform with sub-microsecond temporal resolution. *Nature Methods* **9**, 487-492 (2012).

29 Lin, C.-Y. *et al.* Ultrafast polymer dynamics through a nanopore. *Nano Letters* **22**, 8719-8727 (2022).

30 Chen, K. *et al.* Dynamics of driven polymer transport through a nanopore. *Nature Physics* **17**, 1043-1049 (2021).

31 Ivanov, A. P. *et al.* DNA tunneling detector embedded in a nanopore. *Nano Letters* **11**, 279-285 (2011).

32 Di Ventra, M. & Taniguchi, M. Decoding DNA, RNA and peptides with quantum tunnelling. *Nature Nanotechnology* **11**, 117-126 (2016).

33 Waduge, P. *et al.* Nanopore-based measurements of protein size, fluctuations, and conformational changes. *ACS Nano* **11**, 5706-5716 (2017).

34 Yang, J., Pan, T., Xie, Z., Yuan, W. & Ho, H.-P. In-tube micro-pyramidal silicon nanopore for inertial-kinetic sensing of single molecules. *Nature Communications* **15**, 5132 (2024).
35 Yang, J. *et al.* Angular-Inertia Regulated Stable and Nanoscale Sensing of Single Molecules Using Nanopore-In-A-Tube. *Advanced Materials* **37**, 2400018 (2025).
36 Wanunu, M. *et al.* Rapid electronic detection of probe-specific microRNAs using thin nanopore sensors. *Nature nanotechnology* **5**, 807-814 (2010).
37 Wen, C., Zhang, Z. & Zhang, S.-L. Physical Model for Rapid and Accurate Determination of Nanopore Size via Conductance Measurement. *ACS Sensors* **2**, 1523-1530 (2017).
38 Lastra, L. S., Bandara, Y. M. N. D. Y., Nguyen, M., Farajpour, N. & Freedman, K. J. On the origins of conductive pulse sensing inside a nanopore. *Nature Communications* **13**, 2186 (2022).
39 Zhai, S. & Zhao, H. Influence of concentration polarization on DNA translocation through a nanopore. *Physical Review E* **93**, 052409 (2016).
40 Hautus, M. J., Macmillan, N. A. & Creelman, C. D. *Detection theory: A user's guide*. (Routledge, 2021).
41 Wick, R. R., Judd, L. M. & Holt, K. E. Performance of neural network basecalling tools for Oxford Nanopore sequencing. *Genome biology* **20**, 129 (2019).
42 Koutroumbas, K. & Theodoridis, S. *Pattern recognition*. (Academic Press, 2008).
43 Chen, H. *et al.* Ionic strength-dependent persistence lengths of single-stranded RNA and DNA. *Proceedings of the National Academy of Sciences* **109**, 799-804 (2012).
44 Blackburn, G. M. & Chemistry, R. S. o. *Nucleic Acids in Chemistry and Biology*. (RSC Pub., 2006).
45 Šponer, J., Leszczynski, J. & Hobza, P. Electronic properties, hydrogen bonding, stacking, and cation binding of DNA and RNA bases. *Biopolymers* **61**, 3-31 (2001).
46 Lagerqvist, J., Zwolak, M. & Di Ventra, M. Fast DNA Sequencing via Transverse Electronic Transport. *Nano Letters* **6**, 779-782 (2006).

**Acknowledgment**

We thank Dr. Wing Yin Ng for valuable contribution to the development of the readout framework; This work was supported in part by the Research Grant Council (RGC) of Hong Kong SAR through GRF projects (14208523, 14211223, 14204621).

**Author contributions**

H.-P.H. conceived the idea of the work. J.Y., and D.H. fabricated the pore for the initial studies. H.-P.H. optimized experimental designs and fabrication protocols. J.Y. and T.P. investigated the device assembly. H.-P.H. discussed and improved the research mechanism. J.Y. and T.P. developed the data analysis framework. J.Y. and D.H. performed pore test, signal characterization, and other related experiments. J.Y. and D.H. prepared the manuscript with thorough editing and polishing from W.Y. and H.-P.H.

**Competing interests**

The authors declare no competing financial interests.

**Data availability**

All data needed to evaluate the conclusions in the paper are present in the paper and/or the Supplementary Materials. Additional data related to this paper may be requested from the corresponding authors (H.-P.H).

## Additional information

Supplementary information is available in the online version.